\documentclass[12pt,a4paper]{article}
\usepackage{amssymb,amsmath}
\allowdisplaybreaks
\begin{document}
\title{Finslerian 4-spinors as a generalization of twistors}
\author{A.V. Solov'yov\thanks{Division of Theoretical Physics, Faculty of
  Physics, Moscow State University, Moscow, Russia. E-mail:
  anton@spin.phys.msu.ru}}
\date{}
\maketitle
\begin{abstract}
The main facts of the geometry of Finslerian 4-spinors are formulated. It is
 shown that twistors are a special case of Finslerian 4-spinors. The close
 connection between Finslerian 4-spinors and the geometry of a 16-dimensional
 vector Finslerian space is established. The isometry group of this space is
 described. The procedure of dimensional reduction to 4-dimensional quantities
 is formulated.
\end{abstract}
\section{Introduction}
In the works \cite{Finkelstein:Hyperspin, Finkelstein:Hypermanifolds},
\textit{hyperspinors} and their basic properties were considered. The same
mathematical objects were independently studied under the name of $N$-component
spinors in the papers \cite{Vladimirov:3-spinors, Solov'yov:4-spinors}. Finally,
in the work \cite{Solo_Vlad:N-spinors}, the general algebraic theory of
\textit{Finslerian $N$-spinors} was constructed. The last term is more suitable
because it reflects the close connection between hyperspinors and Finslerian
geometry.

This paper is devoted to formulating the main facts of the geometry of
Finslerian 4-spinors. We show that \textit{twistors} of R.~Penrose
\cite{Penrose} are a special case of Finslerian 4-spinors and can be associated
not only with pseudo-Euclidean geometry, but also with Finslerian one. After
deducing the expression for the length of a vector in the 16-dimensional
Finslerian space, we describe the corresponding isometry group. We also
formulate the procedure of dimensional reduction which allows us to rewrite the
expression for the Finslerian length of a 16-vector in terms of 4-dimensional
geometric objects.
\section{The geometry of Finslerian 4-spinors}
Let $\mathbb{C}^4$ be the vector space of 4-component columns of complex numbers
with respect to the standard matrix operations of addition and multiplication by
elements of the field $\mathbb{C}$. Let us consider the antisymmetric 4-linear
form
\begin{equation}
[\xi,\eta,\lambda,\mu]=\varepsilon_{abcd}\,\xi^a\eta^b\lambda^c\mu^d,
\label{eq:1}
\end{equation}
where $\xi$, $\eta$, $\lambda$, $\mu\in\mathbb{C}^4$, $\varepsilon_{abcd}$ is
the Levi-Civita symbol with the ordinary normalization $\varepsilon_{1234}=1$,
the indices $a$, $b$, $c$, $d$ run independently from 1 to 4, and $\xi^a$,
$\eta^b$, $\lambda^c$, $\mu^d\in\mathbb{C}$. Here and in the following formulas,
the summation is taken over all the repeating indices.

The space $\mathbb{C}^4$ equipped with the form \eqref{eq:1} is called the
\textit{space of Finslerian 4-spinors}. The complex number $[\xi,\eta,\lambda,
\mu]$ is respectively called the \textit{symplectic scalar 4-product} of the
Finslerian 4-spinors $\xi$, $\eta$, $\lambda$, and $\mu$.

Since \eqref{eq:1} is the determinant
\begin{equation}
[\xi,\eta,\lambda,\mu]=
\begin{vmatrix}
\xi^1&\eta^1&\lambda^1&\mu^1\\
\xi^2&\eta^2&\lambda^2&\mu^2\\
\xi^3&\eta^3&\lambda^3&\mu^3\\
\xi^4&\eta^4&\lambda^4&\mu^4
\end{vmatrix}
\label{eq:2}
\end{equation}
with the columns $\xi$, $\eta$, $\lambda$, $\mu$, the symplectic scalar
4-product $[\xi,\eta,\lambda,\mu]$ vanishes if and only if the Finslerian
4-spinors $\xi$, $\eta$, $\lambda$, and $\mu$ are linearly dependent
\cite{Kostrikin}. In particular, $[\xi,\xi,\xi,\xi]=0$ for any
$\xi\in\mathbb{C}^4$.

Let us find isometries of the space of Finslerian 4-spinors, i.e., the linear
transformations
\begin{equation}
\xi^\prime=D\xi\quad\Longleftrightarrow\quad\xi^{\prime a}=d^a_b\xi^b\quad
(D=\|d^a_b\|; d^a_b\in\mathbb{C}; a,b=\overline{1,4})
\label{eq:3}
\end{equation}
which preserve the symplectic scalar 4-product:
\begin{equation}
[\xi^\prime,\eta^\prime,\lambda^\prime,\mu^\prime]=[\xi,\eta,\lambda,\mu]
\quad\text{for any}\quad\xi,\eta,\lambda,\mu\in\mathbb{C}^4.
\label{eq:4}
\end{equation}
Substituting \eqref{eq:3} and the similar expressions for $\eta^\prime$,
$\lambda^\prime$, $\mu^\prime$ into the condition \eqref{eq:4}, we obtain
\begin{equation}
[\xi,\eta,\lambda,\mu]\det D=[\xi,\eta,\lambda,\mu]
\label{eq:5}
\end{equation}
with regard to \eqref{eq:2}. Due to arbitrariness of $\xi$, $\eta$, $\lambda$,
$\mu\in\mathbb{C}^4$, the equation \eqref{eq:5} implies $\det D=1$. Thus, the
isometries of the space of Finslerian 4-spinors form the group
$\text{SL}(4,\mathbb{C})$.

If to equip $\mathbb{C}^4$ with the additional geometric structure, then the
space of Finslerian 4-spinors becomes the twistor space. Namely, let us consider
the Hermitian form
\begin{equation}
\langle\xi,\eta\rangle=\xi^1\overline{\eta^1}+\xi^2\overline{\eta^2}-
\xi^3\overline{\eta^3}-\xi^4\overline{\eta^4},
\label{eq:6}
\end{equation}
where $\xi$, $\eta\in\mathbb{C}^4$ and the over-lines denote complex
conjugating. The complex number $\langle\xi,\eta\rangle$ is usually called the
pseudounitary scalar product of $\xi$ and $\eta$. With respect to the scalar
product \eqref{eq:6}, $\mathbb{C}^4$ is the twistor space~\cite{Penrose}. It is
evident that the transformations \eqref{eq:3}, which preserve the forms
\eqref{eq:1} and \eqref{eq:6} simultaneously, make up the so-called twistor
group $\text{SU}(2,2)\subset\text{SL}(4,\mathbb{C})$. In this sense, twistors
are a special case of Finslerian 4-spinors.

Let us consider the subspace of the vector space $\mathbb{C}^4\otimes
\overline{\mathbb{C}^4}$ which consists of Hermitian tensors. This subspace is
isomorphic to the 16-dimensional \textit{real} vector space $\text{Herm}(4)=
\{X\mid X=X^+\}$ of all Hermitian $4\times 4$ matrices with complex
elements. Here, the cross denotes Hermitian conjugating.

As a basis of the space $\text{Herm}(4)$, we choose the following linearly
independent matrices
\begin{align}
\tau_0&=
\begin{pmatrix}
1&0&0&0\\
0&1&0&0\\
0&0&0&0\\
0&0&0&0
\end{pmatrix},&
\tau_1&=
\begin{pmatrix}
0&1&0&0\\
1&0&0&0\\
0&0&0&0\\
0&0&0&0
\end{pmatrix},&
\tau_2&=
\begin{pmatrix}
0&-i&0&0\\
i&0&0&0\\
0&0&0&0\\
0&0&0&0
\end{pmatrix},\notag\\
\tau_3&=
\begin{pmatrix}
1&0&0&0\\
0&-1&0&0\\
0&0&0&0\\
0&0&0&0
\end{pmatrix},&
\tau_4&=
\begin{pmatrix}
0&0&1&0\\
0&0&0&0\\
1&0&0&0\\
0&0&0&0
\end{pmatrix},&
\tau_5&=
\begin{pmatrix}
0&0&-i&0\\
0&0&0&0\\
i&0&0&0\\
0&0&0&0
\end{pmatrix},\notag\\
\tau_6&=
\begin{pmatrix}
0&0&0&0\\
0&0&1&0\\
0&1&0&0\\
0&0&0&0
\end{pmatrix},&
\tau_7&=
\begin{pmatrix}
0&0&0&0\\
0&0&-i&0\\
0&i&0&0\\
0&0&0&0
\end{pmatrix},&
\tau_8&=
\begin{pmatrix}
0&0&0&0\\
0&0&0&0\\
0&0&1&0\\
0&0&0&0
\end{pmatrix},\notag\\
\tau_9&=
\begin{pmatrix}
0&0&0&1\\
0&0&0&0\\
0&0&0&0\\
1&0&0&0
\end{pmatrix},&
\tau_{10}&=
\begin{pmatrix}
0&0&0&-i\\
0&0&0&0\\
0&0&0&0\\
i&0&0&0
\end{pmatrix},&
\tau_{11}&=
\begin{pmatrix}
0&0&0&0\\
0&0&0&1\\
0&0&0&0\\
0&1&0&0
\end{pmatrix},\notag\\
\tau_{12}&=
\begin{pmatrix}
0&0&0&0\\
0&0&0&-i\\
0&0&0&0\\
0&i&0&0
\end{pmatrix},&
\tau_{13}&=
\begin{pmatrix}
0&0&0&0\\
0&0&0&0\\
0&0&0&1\\
0&0&1&0
\end{pmatrix},&
\tau_{14}&=
\begin{pmatrix}
0&0&0&0\\
0&0&0&0\\
0&0&0&-i\\
0&0&i&0
\end{pmatrix},\notag\\
\tau_{15}&=
\begin{pmatrix}
0&0&0&0\\
0&0&0&0\\
0&0&0&0\\
0&0&0&1
\end{pmatrix}.
\label{eq:7}
\end{align}
Then, for any $X\in\text{Herm}(4)$, we have the expansion 
\begin{equation}
X=X^A\tau_A\quad (A=\overline{0,15}),
\label{eq:8}
\end{equation}
where $X^A\in\mathbb{R}$ are components of the 16-vector $X$ with respect to the
basis \eqref{eq:7}. Along with the matrices \eqref{eq:7}, we introduce another
set of the Hermitian $4\times 4$ matrices: $\tau^B=\tau_B$ ($B\ne8,15$),
$\tau^8=2\tau_8$, $\tau^{15}=2\tau_{15}$. Under such a choice of the matrices,
the remarkable relations
\begin{equation}
\text{Tr}(\tau^A\tau_B)=2\delta^A_B\quad (A,B=\overline{0,15})
\label{eq:9}
\end{equation}
are fulfilled. Here, $\delta^A_B$ is the Kronecker symbol. Because of
\eqref{eq:8} and \eqref{eq:9},
 \begin{equation}
X^A=\frac{1}{2}\text{Tr}(\tau^A X).
\label{eq:10}
\end{equation}

Let us equip $\text{Herm}(4)$ with the structure of the Finslerian space. To
this end, we define the \textit{length} $|X|$ of the 16-vector
$X\in\text{Herm}(4)$ in the following way: $|X|\equiv\sqrt[4]{\det
X}$. Computing the determinant of \eqref{eq:8}, we obtain the expression for
$|X|^4$ in the basis \eqref{eq:7}:
\begin{align}
|X|^4&=G_{ABCD}X^A X^B X^C X^D\notag\\
&=X^{15}\bigl\{[(X^0)^2-(X^1)^2-(X^2)^2-(X^3)^2]X^8\notag\\
&-[(X^4)^2+(X^5)^2+(X^6)^2+(X^7)^2]X^0+2[X^4X^6+X^5X^7]X^1\notag\\
&+2[X^5X^6-X^4X^7]X^2+[(X^4)^2+(X^5)^2-(X^6)^2-(X^7)^2]X^3\bigr\}\notag\\
&-[(X^0)^2-(X^1)^2-(X^2)^2-(X^3)^2][(X^{13})^2+(X^{14})^2]\notag\\
&+[(X^4)^2+(X^5)^2][(X^{11})^2+(X^{12})^2]+[(X^6)^2+(X^7)^2]\notag\\
&\times[(X^9)^2+(X^{10})^2]-X^0X^8[(X^9)^2+(X^{10})^2+(X^{11})^2+(X^{12})^2]\notag\\
&+X^3X^8[(X^9)^2+(X^{10})^2-(X^{11})^2-(X^{12})^2]+2\bigl\{[X^0-X^3]\notag\\
&\times[X^4X^9X^{13}+X^4X^{10}X^{14}-X^5X^9X^{14}+X^5X^{10}X^{13}]\notag\\
&+[X^0+X^3][X^6X^{11}X^{13}+X^6X^{12}X^{14}-X^7X^{11}X^{14}\notag\\
&+X^7X^{12}X^{13}]-X^1[X^4X^{11}X^{13}+X^4X^{12}X^{14}-X^5X^{11}X^{14}\notag\\
&+X^5X^{12}X^{13}+X^6X^9X^{13}+X^6X^{10}X^{14}-X^7X^9X^{14}\notag\\
&+X^7X^{10}X^{13}-X^8X^9X^{11}-X^8X^{10}X^{12}]-X^2[X^4X^{11}X^{14}\notag\\
&-X^4X^{12}X^{13}+X^5X^{11}X^{13}+X^5X^{12}X^{14}-X^6X^9X^{14}\notag\\
&+X^6X^{10}X^{13}-X^7X^9X^{13}-X^7X^{10}X^{14}+X^8X^9X^{12}\notag\\
&-X^8X^{10}X^{11}]-X^4[X^6X^9X^{11}+X^6X^{10}X^{12}+X^7X^9X^{12}\notag\\
&-X^7X^{10}X^{11}]+X^5[X^6X^9X^{12}-X^6X^{10}X^{11}-X^7X^9X^{11}\notag\\
&-X^7X^{10}X^{12}]\bigr\}.
\label{eq:11}
\end{align}
Here, $G_{ABCD}$ are components of the covariant symmetric tensor on
$\text{Herm}(4)$. Thus, the Finslerian length of the 16-vector
$X\in\text{Herm}(4)$ in the basis \eqref{eq:7} is the form of degree 4 with
respect to its components \eqref{eq:10}. It should be noted that the form
\eqref{eq:11} is indefinite, i.e., the cases $|X|^4>0$, $|X|^4<0$ or
$|X|^4=0$ are possible. Since $|X|^4=\det X$, we have $|X|^4=0$ if and only if
$\det X=0$.

Any linear transformation \eqref{eq:3} of the space of Finslerian 4-spinors
induces the transformation
\begin{equation}
X^\prime=DXD^+\quad\Longleftrightarrow\quad
X^{\prime a\dot b}=d^a_c\overline{d^{\dot b}_{\dot e}}X^{c\dot e}\quad
(X^\prime=\|X^{\prime a\dot b}\|; X=\|X^{c\dot e}\|)
\label{eq:12}
\end{equation}
in $\text{Herm}(4)$. Here, all the indices run from 1 to 4 and
$X\in\text{Herm}(4)$. It is evident that the transformation \eqref{eq:12} has
the following properties:
\begin{enumerate}
\item If $X=X^+$, then $X^\prime=X^{\prime+}$.
\item The transformation \eqref{eq:12} is linear with respect to $X$.
\item If $\det D=1$, then $\det X^\prime=\det X$ for any $X\in\text{Herm}(4)$.
\end{enumerate}
Since $|X|=\sqrt[4]{\det X}$, the last property means that the linear
transformation \eqref{eq:12} with $D\in\text{SL}(4,\mathbb{C})$ is a Finslerian
isometry of the space $\text{Herm}(4)$, i.e., $|X^\prime|=|X|$. It is clear that
all such isometries form a group. We will give the explicit matrix description
of this group in the basis \eqref{eq:7}.

Let us substitute the expansions $X^\prime=X^{\prime A}\tau_A$ and $X=X^B\tau_B$
into \eqref{eq:12}. We then multiply the resulting equality by $\tau^A$ from the
left, compute its trace, and use the relations \eqref{eq:9}. As a result, we
obtain
\begin{equation}
X^{\prime A}=L(D)^A_B X^B\quad (A,B=\overline{0,15}),
\label{eq:13}
\end{equation}
where
\begin{equation}
L(D)^A_B=\frac{1}{2}\text{Tr}(\tau^A D\tau_B D^+)
\label{eq:14}
\end{equation}
are elements of the matrix of the linear transformation \eqref{eq:12} in the
basis \eqref{eq:7}. It should be noted that $L(D)^A_B\in\mathbb{R}$. Thus, for
any $D\in\text{SL}(4,\mathbb{C})$, the transformation
\eqref{eq:13}--\eqref{eq:14} preserves the form \eqref{eq:11} so that
$G_{ABCD}X^{\prime A}X^{\prime B}X^{\prime C}X^{\prime D}=G_{ABCD}X^A X^B X^C
X^D$. 

Since the group $\text{SL}(2,\mathbb{C})\subset\text{SL}(4,\mathbb{C})$ is
locally isomorphic to the group $\text{O}^\uparrow_+(1,3)$~\cite{Postnikov}, it
is interesting to consider the transformation \eqref{eq:13}--\eqref{eq:14} with
$D\in\text{SL}(2,\mathbb{C})$, i.e., from the point of view of a ``4-dimensional
observer''. This will allow us to represent the expression \eqref{eq:11} for the
Finslerian length of the 16-vector completely in the 4-dimensional form.

Let
\begin{equation}
D_2=
\begin{pmatrix}
d^1_1&d^1_2&0&0\\
d^2_1&d^2_2&0&0\\
0&0&1&0\\
0&0&0&1
\end{pmatrix},\quad
\det D_2=1\quad
(d^{\hat a}_{\hat b}\in\mathbb{C}; \hat a,\hat b=1,2).
\label{eq:15}
\end{equation}
The matrices \eqref{eq:15} form a subgroup of $\text{SL}(4,\mathbb{C})$ which is
isomorphic to the group $\text{SL}(2,\mathbb{C})$. Let us substitute the matrix
$D_2$ from \eqref{eq:15} into \eqref{eq:14} instead of $D$. Direct computations
show that
\begin{align}
L(D_2)^0_0&=\frac{1}{2}(
d^1_1\overline{d^1_1}+
d^1_2\overline{d^1_2}+
d^2_1\overline{d^2_1}+
d^2_2\overline{d^2_2}),\notag\\
L(D_2)^0_1&=\frac{1}{2}(
d^1_1\overline{d^1_2}+
d^2_1\overline{d^2_2}+
d^1_2\overline{d^1_1}+
d^2_2\overline{d^2_1}),\notag\\
L(D_2)^0_2&=\frac{i}{2}(
d^1_2\overline{d^1_1}+
d^2_2\overline{d^2_1}-
d^1_1\overline{d^1_2}-
d^2_1\overline{d^2_2}),\notag\\
L(D_2)^0_3&=\frac{1}{2}(
d^1_1\overline{d^1_1}+
d^2_1\overline{d^2_1}-
d^1_2\overline{d^1_2}-
d^2_2\overline{d^2_2}),\notag\\
L(D_2)^1_0&=\frac{1}{2}(
d^1_1\overline{d^2_1}+
d^2_1\overline{d^1_1}+
d^1_2\overline{d^2_2}+
d^2_2\overline{d^1_2}),\notag\\
L(D_2)^1_1&=\frac{1}{2}(
d^1_1\overline{d^2_2}+
d^2_1\overline{d^1_2}+
d^1_2\overline{d^2_1}+
d^2_2\overline{d^1_1}),\notag\\
L(D_2)^1_2&=\frac{i}{2}(
d^1_2\overline{d^2_1}+
d^2_2\overline{d^1_1}-
d^1_1\overline{d^2_2}-
d^2_1\overline{d^1_2}),\notag\\
L(D_2)^1_3&=\frac{1}{2}(
d^1_1\overline{d^2_1}+
d^2_1\overline{d^1_1}-
d^1_2\overline{d^2_2}-
d^2_2\overline{d^1_2}),\notag\\
L(D_2)^2_0&=\frac{i}{2}(
d^1_1\overline{d^2_1}-
d^2_1\overline{d^1_1}+
d^1_2\overline{d^2_2}-
d^2_2\overline{d^1_2}),\notag\\
L(D_2)^2_1&=\frac{i}{2}(
d^1_1\overline{d^2_2}-
d^2_1\overline{d^1_2}+
d^1_2\overline{d^2_1}-
d^2_2\overline{d^1_1}),\notag\\
L(D_2)^2_2&=\frac{1}{2}(
d^1_1\overline{d^2_2}+
d^2_2\overline{d^1_1}-
d^1_2\overline{d^2_1}-
d^2_1\overline{d^1_2}),\notag\\
L(D_2)^2_3&=\frac{i}{2}(
d^1_1\overline{d^2_1}-
d^2_1\overline{d^1_1}-
d^1_2\overline{d^2_2}+
d^2_2\overline{d^1_2}),\notag\\
L(D_2)^3_0&=\frac{1}{2}(
d^1_1\overline{d^1_1}-
d^2_1\overline{d^2_1}+
d^1_2\overline{d^1_2}-
d^2_2\overline{d^2_2}),\notag\\
L(D_2)^3_1&=\frac{1}{2}(
d^1_1\overline{d^1_2}-
d^2_1\overline{d^2_2}+
d^1_2\overline{d^1_1}-
d^2_2\overline{d^2_1}),\notag\\
L(D_2)^3_2&=\frac{i}{2}(
d^1_2\overline{d^1_1}-
d^2_2\overline{d^2_1}-
d^1_1\overline{d^1_2}+
d^2_1\overline{d^2_2}),\notag\\
L(D_2)^3_3&=\frac{1}{2}(
d^1_1\overline{d^1_1}-
d^1_2\overline{d^1_2}-
d^2_1\overline{d^2_1}+
d^2_2\overline{d^2_2}),
\label{eq:16}
\end{align}
$L(D_2)^{3+i}_{3+j}=L(D_2)^{8+i}_{8+j}=M(D_2)^i_j$ ($i,j=\overline{1,4}$), where
\begin{align}
M(D_2)^1_1&=\frac{1}{2}(\overline{d^1_1}+d^1_1),\quad
M(D_2)^3_1=\frac{1}{2}(\overline{d^2_1}+d^2_1),\notag\\
M(D_2)^1_2&=\frac{i}{2}(\overline{d^1_1}-d^1_1),\quad
M(D_2)^3_2=\frac{i}{2}(\overline{d^2_1}-d^2_1),\notag\\
M(D_2)^1_3&=\frac{1}{2}(\overline{d^1_2}+d^1_2),\quad
M(D_2)^3_3=\frac{1}{2}(\overline{d^2_2}+d^2_2),\notag\\
M(D_2)^1_4&=\frac{i}{2}(\overline{d^1_2}-d^1_2),\quad
M(D_2)^3_4=\frac{i}{2}(\overline{d^2_2}-d^2_2),\notag\\
M(D_2)^2_1&=\frac{i}{2}(d^1_1-\overline{d^1_1}),\quad
M(D_2)^4_1=\frac{i}{2}(d^2_1-\overline{d^2_1}),\notag\\
M(D_2)^2_2&=\frac{1}{2}(d^1_1+\overline{d^1_1}),\quad
M(D_2)^4_2=\frac{1}{2}(d^2_1+\overline{d^2_1}),\notag\\
M(D_2)^2_3&=\frac{i}{2}(d^1_2-\overline{d^1_2}),\quad
M(D_2)^4_3=\frac{i}{2}(d^2_2-\overline{d^2_2}),\notag\\
M(D_2)^2_4&=\frac{1}{2}(d^1_2+\overline{d^1_2}),\quad
M(D_2)^4_4=\frac{1}{2}(d^2_2+\overline{d^2_2}),
\label{eq:17}
\end{align}
$L(D_2)^{8}_{8}=L(D_2)^{13}_{13}=L(D_2)^{14}_{14}=L(D_2)^{15}_{15}=1$, while the
other elements of the matrix of the transformation $X^{\prime A}=L(D_2)^A_B X^B$
vanish. Thus, for $D=D_2$, the  Finslerian isometry \eqref{eq:13} has the form
\begin{align}
X^{\prime\alpha}&=L(D_2)^\alpha_\beta X^\beta\quad
(\alpha,\beta=\overline{0,3}),\notag\\
\theta^{\prime i}&=M(D_2)^i_j\theta^j\quad
(i,j=\overline{1,4}),\notag\\
X^{\prime 8}&=X^8,\notag\\
\vartheta^{\prime i}&=M(D_2)^i_j\vartheta^j\quad
(i,j=\overline{1,4}),\notag\\
X^{\prime 13}&=X^{13},\notag\\
X^{\prime 14}&=X^{14},\notag\\
X^{\prime 15}&=X^{15},
\label{eq:18}
\end{align}
where $L(D_2)^\alpha_\beta$, $M(D_2)^i_j$ are given by
\eqref{eq:16}--\eqref{eq:17} and the notation $\theta^i=X^{3+i}$,
$\vartheta^j=X^{8+j}$ is used.

It was shown in the paper \cite{Solo_Vlad:N-spinors} that  \eqref{eq:16} and
\eqref{eq:17} are the elements of the matrices of the transformations for a
Lorentz 4-vector and a Majorana 4-spinor respectively. Therefore, the result
\eqref{eq:18} asserts that, for $D=D_2$, the 16-vector $X^A$ splits into the
Lorentz 4-vector  $X^\alpha$, the Majorana 4-spinors $\theta^i$, $\vartheta^j$,
and the Lorentz 4-scalars $X^8$, $X^{13}$, $X^{14}$, $X^{15}$.

This is the essence of the procedure of dimensional reduction allowing to
display the ``4-dimensional structure'' of 16-dimensional expressions. Let us
apply this procedure to the cumbersome formula \eqref{eq:11} for the Finslerian
length of the 16-vector $X^A$. Taking into consideration \eqref{eq:18}, we
obtain
\begin{align}
|X|^4&=X^{15}[X^8g_{\mu\nu}X^\mu X^\nu-g_{\mu\nu}X^\mu\overline{\theta}
\gamma^\nu\theta]\notag\\
&-[(X^{13})^2+(X^{14})^2]g_{\mu\nu}X^\mu X^\nu-X^8g_{\mu\nu}X^\mu
\overline{\vartheta}\gamma^\nu\vartheta\notag\\
&+2X^{13}g_{\mu\nu}X^\mu\overline{\theta}\gamma^\nu\vartheta+2X^{14}g_{\mu\nu}
X^\mu\overline{\theta}\gamma^5\gamma^\nu\vartheta\notag\\
&+\frac{1}{2}g_{\mu\nu}\overline{\theta}\gamma^\mu\theta\,
\overline{\vartheta}\gamma^\nu\vartheta,
\label{eq:19}
\end{align}
where $\mu,\nu=\overline{0,3}$, $\|g_{\mu\nu}\|=\text{diag}\,(1,-1,-1,-1)$ is
the matrix of components of the Minkowski metric tensor in a pseudoorthonormal
basis,
\begin{align*}
&\gamma^0=
\begin{pmatrix}
0&0&i&0\\
0&0&0&-i\\
-i&0&0&0\\
0&i&0&0
\end{pmatrix},\
\gamma^1=
\begin{pmatrix}
i&0&0&0\\
0&-i&0&0\\
0&0&-i&0\\
0&0&0&i
\end{pmatrix},\
\gamma^2=
\begin{pmatrix}
0&i&0&0\\
i&0&0&0\\
0&0&0&i\\
0&0&i&0
\end{pmatrix},\\
&\gamma^3=
\begin{pmatrix}
0&0&-i&0\\
0&0&0&i\\
-i&0&0&0\\
0&i&0&0
\end{pmatrix},\
\gamma^5=\gamma^0\gamma^1\gamma^2\gamma^3=
\begin{pmatrix}
0&-1&0&0\\
1&0&0&0\\
0&0&0&-1\\
0&0&1&0
\end{pmatrix}
\end{align*}
are the Dirac matrices in the Majorana representation
\cite{Solo_Vlad:N-spinors}, $\theta,\vartheta\in\mathbb{R}^4$ are the Majorana
4-spinors, and $\overline{\theta}=\theta^\top\gamma^0$, $\overline{\vartheta}=
\vartheta^\top\gamma^0$ (the mark ${}^\top$ denotes the matrix
transposition). Thus, the expression \eqref{eq:11} is written in the compact
4-dimensional form \eqref{eq:19}.
\section{Conclusion}
Summarizing, we make some remarks concerning the obtained results.

First of all, we should note the dual nature of twistors. Those are spinors of
the 6-dimensional pseudo-Euclidean space with two time-like
dimensions~\cite{Penrose}. On the other hand, as it is shown in this paper,
twistors are a special case of Finslerian 4-spinors of the 16-dimensional vector
space equipped with the metric form \eqref{eq:11}.

In addition, the paper contains the explicit description of isometries of the
above 16-dimensional Finslerian space and the procedure of dimensional reduction
which allows us to write \eqref{eq:11} in the 4-dimensional form
\eqref{eq:19}. The latter is important because it demonstrates the
correspondence of our constructions to the standard relativistic theory on the
level of geometry.

The author is grateful to Yu.S.~Vladimirov,  S.V.~Bolokhov, and A.V.~Pili\-penko
for helpful discussions of obtained results.


\begin{thebibliography}{9}
\bibitem{Finkelstein:Hyperspin}D. Finkelstein. \textit{Hyperspin and
  hyperspace}. Physical Review Letters \textbf{56}, 1532--1533 (1986).
\bibitem{Finkelstein:Hypermanifolds}D. Finkelstein, S. R. Finkelstein, and
  C. Holm. \textit{Hyperspin manifolds}. International Journal of Theoretical
  Physics \textbf{25}, 441--463 (1986).
\bibitem{Vladimirov:3-spinors}Yu. S. Vladimirov and A. V. Solov'yov. \textit{The
  physical structure of the rank $(4,4;b)$ and three-component
  spinors}. Novosibirsk: Institute of Mathematics, Sib.\ Otd.\ Akad.\ Nauk SSSR,
  1990. Vychislitel'nye Sistemy, vyp.~135, pp.~44--66 (in Russian).
\bibitem{Solov'yov:4-spinors}A. V. Solov'yov. \textit{On the theory of binary
  physical structures of the rank $(5,5;b)$ and higher}. Novosibirsk: Institute
  of Mathematics, Sib.\ Otd.\ Akad.\ Nauk SSSR, 1990. Vychislitel'nye Sistemy,
  vyp.~135, pp.~67--77 (in Russian).
\bibitem{Solo_Vlad:N-spinors}A. V. Solov'yov and
  Yu. S. Vladimirov. \textit{Finslerian $N$-spinors: Algebra}. International
  Journal of Theoretical Physics \textbf{40}, 1511--1523 (2001).
\bibitem{Penrose}R. Penrose and W. Rindler. Spinors and space-time. Spinor and
  twistor methods in space-time geometry. Cambridge: Cambridge University Press,
  1986.
\bibitem{Kostrikin}A. I. Kostrikin. Introduction to algebra. New York:
  Springer-Verlag, 1982.
\bibitem{Postnikov}M. M. Postnikov. Lectures on geometry. Linear
  algebra. Moscow: Nauka, 1986 (in Russian).
\end{thebibliography}
\end{document}